\documentclass[12pt,preprint]{article}

\usepackage{amssymb}
\usepackage{amsmath}
\usepackage{graphics}
\usepackage{epsfig}
\newcommand{\eqb}{\begin{equation}}
\newcommand{\eqe}{\end{equation}}
\newcommand{\dmb}{\begin{displaymath}}
\newcommand{\dme}{\end{displaymath}}

\newcommand{\eab}{\begin{eqnarray}}
\newcommand{\eae}{\end{eqnarray}}

\newcommand{\e}{\mbox{e}}
\newcommand{\be}{\begin{equation}}
\newcommand{\ee}{\end{equation}}

\begin{document}
\begin{titlepage}

\begin{center}
\Large{Low-frequency line temperatures of the CMB}\vspace{2.5cm}\\ 
\large{Ralf Hofmann}
\end{center}
\vspace{1.5cm} 
\begin{center}
{\em Institut f\"ur Theoretische Physik\\ 
Universit\"at Heidelberg\\ 
Philosohenweg 16\\ 
69120 Heidelberg, Germany}
\end{center}
\vspace{5.0cm}
\begin{abstract}
Based on SU(2) Yang-Mills thermodynamics 
we interprete Aracde2's and the results of earlier radio-surveys on low-frequency CMB line temperatures 
as a phase-boundary effect. We explain the excess at low 
frequencies by evanescent, nonthermal 
photon fields of the CMB whose intensity is nulled by that of Planck distributed calibrator 
photons. The CMB baseline temperature thus is identified with the critical temperature of the 
deconfining-preconfining transition.
\end{abstract} 
\end{titlepage}

\noindent {\sl Introduction.}
Activities to detect deviations of the CMB spectrum from an ideal 
black-body shape and to extract angular correlation functions 
from carefully generated CMB maps are numerous and insightful 
\cite{FIRAS,WMAP}. In particular, the observational situation at 
low frequencies \cite{arcade2,arcade2I,Roger,Maeda,Haslam,Reich} and 
large-angles \cite{WMAP,Schwarz}, respectively, has generated genuine 
surprises. We are convinced that 
these anomalies necessitate changing our present theoretical 
concept on photon physics \cite{SH2007}. Specifically, we mean a replacement of the gauge group 
U(1) by SU(2), the latter being treated nonperturbatively \cite{H2005,SHG2006I,SHG2006II,LH2008}.
This Letter intends to spell out a topical experimental reason confirming this. 
Recent data on CMB line temperatures $T$ at 
low-frequencies ($\nu=3\,\mbox{GHz}\, ...\, 90\,\mbox{GHz}$) \cite{arcade2}, determined 
by nulling the difference between CMB (cleared of galactic emission) and black-body calibrator 
spectral intensities, indicate a statistically significant ($5\,\sigma$) 
excess at the lowest frequencies. Combining this with earlier radio-frequency data 
on forground subtracted antenna temperatures \cite{Roger,Maeda,Haslam,Reich}, a fit to an affine power law
\eqb
\label{lineT}
T(\nu)=T_0+T_R\,\left(\frac{\nu}{\nu_0}\right)^\beta\,
\eqe
reveals \cite{arcade2}: $T_0=2.725\,$K (within errors FIRAS' CMB baseline 
temperature \cite{FIRAS} obtained by a fit to the CMB spectrum at 
high frequencies), $\nu_0=1\,$GHz, $T_R=1.19\pm 0.14\,$K, and a spectral index of $\beta=-2.62\pm 0.04$. 
Arcade2's claim that this spectacular deviation from a perfect black-body situation ($T(\nu)\equiv \mbox{const}$) 
is not an artefact of 
galactic foreground subtraction, unlikely is related to an average effect of 
distant point sources, and that these results naturally continue 
earlier radio-frequency data \cite{Roger,Maeda,Haslam,Reich} convinces 
in light of their arguments. 
The observational situation thus calls for an unconventional explanation of Eq.\,(\ref{lineT}). 
We work in units where $k_B=c=\hbar=1$. In these units the CMB baseline temperature 
assumes the value 356.76 (56.78)\,GHz of a circular (ordinary) frequency.\\ 
 
\noindent {\sl Physics at the phase boundary.}
In the preconfining phase of SU(2) Yang-Mills thermodynamics 
the photon acquires a Meissner mass $m_\gamma=g|\varphi|$ where $g$ is the 
dual gauge coupling which vanishes at $T=T_c$ and 
rises rapidly (critical exponent $\frac{1}{2}$) when $T$ falls below 
$T_c$ \cite{H2005}. Moreover, the modulus $|\varphi|=\sqrt{\frac{\Lambda^3_M}{2\pi T}}$ 
is part of the description of the monopole condensate parameterized 
by the preconfining manifestation $\Lambda_M$ of the Yang-Mills scale. 
On large spatial scales, the superconducting, preconfining 
ground state enforcing this Meissner mass may be responsible 
for the ermergence of extragalactic magnetic fields of thus far unexplained origin.   

It is important to stress that 
$m_\gamma$ is induced and calculable 
in a situation of thermal equilibrium ($T<T_c$) and that it vanishes 
in the deconfining phase, where modulo mild (anti)screening effects peaking at a 
temperature $T\sim 2\,T_c$ and rapidly decaying for larger temperatures, the 
photon is precisely massless. This reflects the fact that a subgroup U(1) of the underlying 
SU(2) gauge symmetry is respected by the deconfining ground state \cite{H2005}.    

The fact that $m_\gamma$ is a 
Meissner mass implies the evanescence of photons of frequency $\omega<m_\gamma$. This, however, 
is {\sl not} what happens in the deconfining phase \cite{SHG2006I,LH2008}. There, by a coupling 
to effective, massive vector modes, the prohibition of photon propagation at low 
temperatures and frequencies \cite{SHG2006I} 
is energetically balanced by the creation of nonrelativistic and charged particles 
(isolated and screened monopoles and antimonopoles \cite{LKGH2008}). 
As a consequence, in the deconfining phase energy leaves the photon sector 
to re-appear in terms of (anti)monopole mass, and no evanescent photon 
fields are generated at frequencies smaller than the square root of the screening function. 
If the temperature precisely matches $T_c$, however, then deconfining SU(2) Yang-Mills thermodynamics 
predicts the absence of any spectral distortions compared to the conventional Planck spectrum of 
photon intensity.

On the preconfining side of the phase boundary Meissner massive photons of circular frequency 
$\omega$ below $m_\gamma$ do not propagate and create a spectral intensity attributed to an oscillating 
{\sl evanescent} photon field which no longer is thermalized. Evanescent `photons' collectively carry the energy 
density $\Delta\rho(T_c)$ that formerly massless CMB photons have lost due to their interaction with 
the new ground state (superconductor \cite{H2005}). Due to their nonpropagating nature frequencies belonging to the 
evanescent, nonthermal, and random photon field are distributed according 
to a Gaussian of width $m_\gamma$ and normalized to $\Delta\rho(T_c)$. 
Since propagating, preconfining-phase photons can genuinely maintain an additional polarization only if their 
frequency is sizeably lower than $\Delta T=T_c-T\ll T_c$ \footnote{This never happens because of the large slope modulus 
of the function $g(T)$ \cite{H2005}.}, we approximately have 
\eqb
\label{deltarho}
\Delta\rho(T_c)=\int_0^\infty d\omega\, \left.\left(I_{\gamma,\tiny\mbox{dec}}-I_{\gamma,\tiny\mbox{prec}}\right)\right|_{T=T_c}\,,\eqe
where 
\eqb
\label{defI}
I_{\gamma,\tiny\mbox{dec}}=\frac{1}{\pi^2}\,\frac{\omega^3}{\exp(\omega/T)-1}\,\ \ \  
\mbox{and}\ \ \ \ 
I_{\gamma,\tiny\mbox{prec}}=\frac{1}{\pi^2}\,\frac{\sqrt{\omega^2-m_\gamma^2}\,\omega^2}{\exp(\omega/T)-1}\,\theta(\omega-m_\gamma)\,.
\eqe
Here $\theta(x)$ is the Heaviside step function: $\theta(x)=0$ for $x<0$, $\theta(x)=1/2$ for $x=0$, 
and $\theta(x)=1$ for $x>0$. Introducing the dimensionless photon 
mass $\mu_\gamma\equiv\frac{m_\gamma}{T_c}$ yields
\eqb
\label{Delta}
\Delta\rho=\frac{T_c^4}{\pi^2}\left(\frac{\mu_\gamma^3}{3}+F(\mu_\gamma)\right)\ \ 
\mbox{where}\ \ F(\mu_\gamma)\equiv\int_{\mu_\gamma}^\infty dy\,\frac{y^2}{\e^y-1}(y-\sqrt{y^2-\mu_\gamma^2})\,.
\eqe
For the CMB spectral intensity, we thus have  
\eqb
\label{IPm>0}
I_{\gamma,\tiny\mbox{prec}}=2\,\frac{\Delta\rho}{\sqrt{2\pi}\,m_\gamma}\,\exp\left(-\frac{\omega^2}{2 m_\gamma^2}\right)+
\theta(\omega-m_\gamma)\frac{1}{\pi^2}\,\frac{\sqrt{\omega^2-m_\gamma^2}\omega^2}{\exp(\omega/T_c)-1}\,.
\eqe
Since $\omega/T_c\ll 1$ (with $\nu\le 3.4\,$GHz we have for circular frequencies: 
$\omega\le 21.5\,$GHz; and for line temperatures (units of circular frequency): $T\ge T_c=356\,$GHz) 
we are deep inside the Rayleigh-Jeans regime, and thus for 
calibrator photons, which are precisely massless, see below, we may 
write
\eqb
\label{decmassless}
I_{\gamma,\tiny\mbox{dec}}=\frac{\omega^2 T}{\pi^2}\,.
\eqe
Let us again explain the physics underlying Eqs.\,(\ref{decmassless}) and (\ref{IPm>0}). Assume that the CMB temperature 
is just slightly below $T_c$. This introduces a tiny coupling to the SU(2) preconfining ground 
state which endows low-frequency photons with a Meissner mass $m_\gamma$ if they have 
propagated for a sufficiently long time above this ground state whose correlation length at 
$T_c$ is of the order of 1\,km \cite{H2005}. This is certainly true for CMB photons. As a consequence, 
modes with $\omega<m_\gamma$ become evanescent, thus nonthermal, and are spectrally distributed in frequency according to the first 
term in Eq.\,(\ref{IPm>0}). 
For $\omega>m_\gamma$ CMB photons do propagate albeit with a suppression in intensity as compared 
to the ideal Planck spectrum. In principle, some should propagate with three polarizations. 
Due to a mode's increasing ignorance towards the existence of a Meissner-mass-inducing ground state 
this will on average relax to two polarizations for $\omega\gg m_\gamma$. Therefore, 
the spectral model of Eq.\,(\ref{IPm>0}) is not to 
be taken literally for small, {\sl propagating} frequencies although the according 
spectral {\sl integral} is.    

A calibrator photon, on the other hand, is fresh in that the distance between emission at 
the black-body wall and absorption at the radiometer is just a small multiple of its wave 
length. For sufficiently small coupling $g$ (or for $T$ sufficiently close to but below $T_c$) this 
short propagation path is therefore insufficient to generate a mass $m_\gamma$ even at low frequencies. 
As a consequence, none of the calibrator modes is forced into evanescence. To summarize: 
CMB frequencies approximately obey the spectral distribution $I_{\gamma,\tiny\mbox{prec}}$, see Eq.\,(\ref{IPm>0}), 
while low-frequency calibrator photons are distributed according to $I_{\gamma,\tiny\mbox{dec}}$, see Eq.\,(\ref{decmassless}). 
From now on we set $T_c$ equal to the CMB baseline temperature (expressed in terms of a circular frequency): 
$T_c=356.76\,$GHz.\\ 

\begin{figure}
\begin{center}
\leavevmode
\leavevmode
\vspace{4.5cm}
\includegraphics{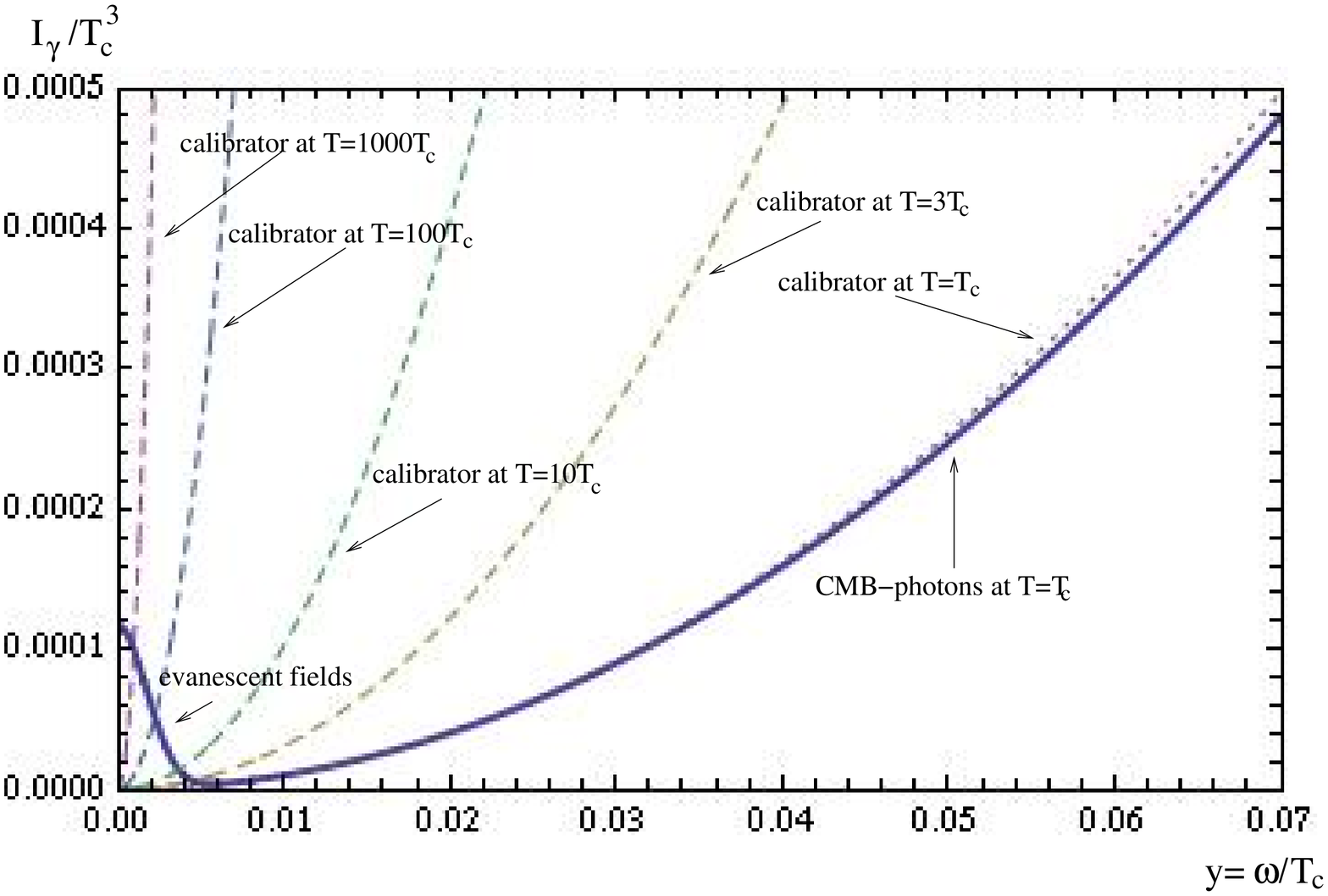}
\end{center}
\caption{The normalized spectral intensities of CMB modes (thick line) at $T=T_c$ and 
$m_\gamma=0.1\,$GHz (in units of ordinary frequency) and 
of calibrator modes (dotted and dashed lines) at various temperatures. A null experiment asks 
for an intersection of the former with a representative of the one-parameter ($T$) familiy of the 
latter-type curves at a given frequency. Since for 
$y\to 0$ the Gaussian becomes stationary one has in this limit $T(y)=\mbox{const}\times y^{-2}$, that is, 
the asymptotic spectral index reads $\beta_{\tiny\mbox{as}}=-2$. With the low-frequency 
data presently available 
one has $\beta\sim -2.6$ \cite{arcade2}.  
\label{Fig-1}}      
\end{figure}

\noindent {\sl Determination of $m_\gamma$ from radio-frequency survey data.} 
The essence of Aracde2's and earlier radio-frequency survey's experimental philosophy 
is to null at a given frequency the CMB intensity signal by that of a calibrator 
black body or of an internal reference load. (Notice that at the low frequencies 
considered there is practically no difference between antenna and thermodynamical temperature \cite{arcade2}.) 
Thus the observationally imposed condition for the extraction of a line temperature $T(\nu)$ is: 
\eqb
\label{null}
I_{\gamma,\tiny\mbox{prec}} = I_{\gamma,\tiny\mbox{dec}}\,.
\eqe
Assuming $m_\gamma=0.1\,$GHz, 
the according spectral situation is depicted in Fig.\,\ref{Fig-1}.
For the extraction of $m_\gamma$ from the data let us 
introduce the following two dimensionless 
quantities
\eqb
\label{dimlessqu}
y\equiv\frac{\omega}{T_c}\,,\ \ \ \ \ \ \ \ \ \ \tau\equiv\frac{T}{T_c}\,. 
\eqe
With these definitions and appealing to Eqs.\,(\ref{decmassless}), (\ref{IPm>0}), and (\ref{Delta}), 
Eq.\,(\ref{null}) is recast as
\eqb
\label{nullrec}
\tau=\sqrt{\frac{2}{\pi}}\,y^{-2}\,\exp\left(-\frac{y^2}{2\mu^2_\gamma}\right)
\left(\frac{\mu_\gamma^2}{3}+\frac{F(\mu_\gamma)}{\mu_\gamma}\right)+
\theta(y-\mu_\gamma)\frac{\sqrt{y^2-\mu_\gamma^2}}{\e^y-1}\,.
\eqe
The following table lists our results for $m_\gamma$, as extracted from the data using Eq.\,(\ref{nullrec}), in units of ordinary (not circular) 
frequency $\nu$:
\dmb
\label{tabel}
\begin{array}{ccccc}
\mbox{source} & \nu [\mbox{GHz}] & T [\mbox{K}] & \mu_\gamma & m_\gamma [\mbox{GHz}]\\ 
\hline 
\mbox{Roger} & 0.022 & 21200\pm 5125 & 0.001821^{+0.000423}_{-0.000419} & 0.1034^{+0.0240}_{-0.0238}\\ 
\mbox{Maeda} & 0.045 & 4355\pm 520 & 0.001704^{+0.000169}_{-0.000166} & 
0.0968^{+0.0095}_{-0.0095}\\ 
\mbox{Haslam} & 0.408 & 16.24\pm 3.4 & 0.003611^{+0.000152}_{-0.000325} & 0.205^{+0.0086}_{-0.0185}\\ 
\mbox{Reich} & 1.42 & 3.213\pm 0.53 & 0.0093^{+0.0007}_{-0.00153}& 0.528^{+0.0397}_{-0.0869}\\ 
\mbox{Arcade2} & 3.20 & 2.792\pm 0.010 & 0.0211^{+0.0001}_{-0.0001}& 1.198^{+0.0057}_{-0.0057}\\ 
\mbox{Arcade2} & 3.41 & 2.771\pm 0.009 & 0.02253^{+0.0001}_{-0.0001} & 1.279^{+0.0057}_{-0.0057} 
\end{array}
\dme
Notice the good agreement of $m_\gamma$ as extracted from the data of Roger \cite{Roger} and Maeda \cite{Maeda} 
where $\nu<m_\gamma$.  The other data of Haslam \cite{Haslam}, Reich \cite{Reich}, and Arcade2 \cite{arcade2} 
yield $\nu>m_\gamma$ 
which is in the regime where we do 
not expect the spectral model for CMB photons to be 
good (average number of polarizations 
depends nontrivially on frequency). Still, the value of $m_\gamma$ obtained from Haslam's 
data \cite{Haslam} is only twice as large as that arising from the data of Roger \cite{Roger} or 
Maeda \cite{Maeda} at a frequency which is, respectively, twenty or ten times larger!  \\ 
\\ 
\noindent {\sl Meissner mass of $\sim 100\,$MHz?} 
At this point it surely is worthwhile to discuss what it really means 
to have the thermalized photon field (at a temperature $T_0=2.727\,$K) acquire a Meissner 
mass? Is this scenario not ruled out by experiments such as radar vs. laser ranging to the moon and 
the limits on the photon mass obtained by terrestial Coulomb-law measurements or the measurement of the  
magnetic fields of astrophysical objects, see \cite{PDG}. 
The answer is no for the following reason: Whether or not the propagation of the photon is 
altered as compared to conventional wisdom sensitively depends of the 
temperature of the thermal ensemble it belongs to and on its frequency. 
To be above the thermal noise of the CMB any experiment trying to detect a photon mass 
(either directly by looking for deviations in electrostatic or magnetostatic 
field configurations or indirectly by searching for modified dispersion laws 
in propagating photon fields) must work with local energy densities attributed to 
the photon field that are by many orders of magnitude larger than that of the 
CMB\footnote{The existence of a correlation between an electric potential gradient 
and a temperature gradient in 
solid-state systems is known for a long time (thermoelectric power). It is conceivable 
that the SU(2) ground state with its abundance of short-lived charge carriers 
acts as a medium which implies a similar correlation.}. 
Even though a static background field or laser emission or radar does not describe a 
homogeneous thermodynamical setting one may for a rough argument appeal to an adiabatic approximation 
setting the experimental energy density equal to that of thermal 
(deconfining) SU(2)$_{\tiny\mbox{CMB}}$ to deduce the local temperature this energy 
density would correspond to were the experimental system actually thermalized. 
In any experimental circumstance searching for a 
universal (by assumption not dependent on temperature) 
photon mass this would yield a temperature far above $T_0=2.725\,$K. But we 
have shown in \cite{LH2008} how rapidly the thermalized SU(2) photon approaches U(1) behavior 
with increasing temperature by a power-like decrease of the modulus of 
its screening function. For example, the spectral gap $\omega^*/T$ in black-body spectra, 
defining the center of the spectral region where nonabelian effects are most 
pronounced (they decay exponentially for $\omega>\omega^*$) decays as 
$T^{-3/2}$. Thus systems that so far were used to obtain photon-mass 
bounds roughly would correspond to temperatures where the photon behaves in a purely abelian 
way explaining the very low mass bounds obtained. That is, for the 
photon to exhibit measurable deviations in its dispersion law it must belong to a {\sl thermal} bath at 
temperatures from just below $T_0$ (Meissner mass) up to $10\,$K (momentum dependent screening mass), 
say. 

What about the physics just around $T_0$? Is there a possibility that thermodynamics is not 
honoured? For example consider the following set-up. Two blackbodies (BBs), one
at $T_1$ just below $T_0$, the other at
$T_2$ just above $T_0$, are immersed into a photon bath exactly at temperature $T_0$. 
Photons exchanged by the two BBs are restricted to frequencies below 
$m_\gamma\sim 100\,$MHz. Would then not BB1 transfer energy to BB2 due to its larger spectral 
intensity below $m_\gamma$ -- in contradiction to the second law of thermodynamics? 
The answer is no because the BB1 photons supporting this bump in the spectrum 
are evanescent and so, by definition, cannot propagate out of BB1's cavity. Also, if $T_0<T_1<T_2$, 
and both $T_1$ and $T_2$ not too far above $T_0$ then
the rapidly rising with temperature spectral intensity
(in the Rayleigh-Jeans regime linearly) would assure,
as it should, that BB1 warms up at the expense of BB2 despite the small
spectral modifications (screening and antiscreening) at temperatures not far above $T_0$. 
\\ 
\noindent {\sl Discussion and Conclusions.} 
Since we may not trust our spectral model for CMB modes locally if 
$\nu>m_\gamma$ (both expressed as ordinary frequencies) it is not surprising that 
considerable deviations occur for the extracted values of $m_\gamma$ 
compared to the low-frequency situation. The {\sl integral} of the spectral model, which 
enters into the normalization $\Delta\rho$
of half the Gaussian in Eq.\,(\ref{IPm>0}), however, is a quantity that is robust against local changes of the 
spectrum. Thus we are inclined to trust our result $m_\gamma\sim 0.1\,$GHz extracted at low 
frequencies (Roger, Maedan). Based on the present work two predictions, arising from an SU(2) Yang-Mills 
theory being responsible for photon propagation \cite{H2005}, can be made: First, since the low-frequency data 
on line temperatures are efficiently explained by this theory being at its deconfining-preconfining phase boundary one 
has $T_c=2.725\,$K. This allows for a precise prediction of a sizable anomaly 
in the low-frequency part of the thermal 
spectral intensity at higher, absolutely given temperatures, 
say at $T=2\,T_c\sim 5.4\,$K \cite{SHG2006II,LH2008}. 
Second, we predict that the spectral index $\beta$ for the line temperature $T(\nu)$, 
measured by nulling the CMB signal by a black-body 
reference load, approaches $\beta_{\tiny\mbox{as}}=-2$ for $\nu\searrow 0$. The tendency of increase of 
$\beta$ when fitting Eq.\,(\ref{lineT}) to low-frequency as compared to intermediate frequency weighted data 
sets is nicely seen in Tab.\,5 of \cite{arcade2}. 

The here presented strong indication that the CMB is on the verge of undergoing a phase 
transition towards superconductivity at its present baseline 
temperature $T_{\tiny\mbox{CMB}}=2.725\,$K implies radical consequences for 
particle physics \cite{SHG2006II}. Since this process occurs on a time scale of $\sim 2\,$ 
billion years \cite{GH2005} there is no immediate consequence for any form 
of energy consuming life.    

\section*{Acknowledgments}

We would like to acknowledge useful discussions with Markus Schwarz who also 
performed an independent numerical calculation extracting $m_\gamma$ 
from the data. The author is grateful to Josef Ludescher for directing his attention to 
Ref.\,\cite{arcade2} and to a Referee insisting on an extended explanation why there is no 
contradiction with existing experimental photon-mass bounds.


\begin{thebibliography}{99}

\bibitem{FIRAS}
D. J. Fixsen et al., Astrophys. J. {\bf 420}, 457 (1994).\\ 
J. C. Mather et al., Astrophys. J. {\bf 420}, 439 (1994).\\ 
J. C. Mather et al., Astrophys. J. Suppl. {\bf 170} 288 (2007).\\  
G. Hinshaw et al., astro-ph/0603451. 

\bibitem{WMAP}
A. Kogut et al., Astrophys. J. Suppl. {\bf 148}, 161 (2003).\\ 
D. N. Spergel et al., Astrophys. J. Suppl. {\bf 148}, 175 (2003).\\ 
D. N. Spergel et al., astro-ph/0603449.\\ 
L. Page et al., astro-ph/0603450.\\ 
G. Hinshaw et al., astro-ph/0603451.\\ 
N. Jarosik et al., astro-ph/0603452. 

\bibitem{arcade2}
D.J. Fixsen et al., arXiv:0901.0555.

\bibitem{arcade2I}
M. Seiffert et al, arXiv:0901.0559. 

\bibitem{Roger}
R.S. Roger et al., A \& AS, 137, 7 (1999).  

\bibitem{Maeda}
Maeda et al., A \& AS, 140, 145 (1999).  

\bibitem{Haslam}
C.G.T. Haslam et al., A \& A 100, 209 (1981).

\bibitem{Reich}
Reich and Reich, A \& AS, 63, 205 (1986). 

\bibitem{Schwarz}
A. de Oliveira-Costa, M. Tegmark, M. Zaldarriga, and A. Hamilton, Phys. Rev. D {\bf 69} 
063516 (2004). [arXiv:astro-ph/0307282] \\ 
D. J. Schwarz, G. D. Starkman, D. Huterer, and C. J. Copi, Phys. Rev. Lett. {\bf 93}, 
221301 (2004). [arXiv:astro-ph/0403353]\\ 
C. J. Copi, D. Huterer, D. J. Schwarz, and G. D. Starkman, 
Phys. Rev. D {\bf 75}, 023507 (2007).  [astro-ph/0605135]\\
C. J. Copi, D. Huterer, D. J. Schwarz, and G. D. Starkman, arXiv:0808.3767 [astro-ph].\\  
P. Bielewicz, K. M. Gorski, and A. J. Banday, Mon. Not. Roy. Astron. Soc. {\bf 355}, 1283 (2004). [arXiv:astro-ph/0405007] 

\bibitem{SH2007}
M. Szopa and R. Hofmann, JCAP {\bf 0803}:001 (2008). [arXiv:hep-ph/0703119]

\bibitem{H2005} 
R. Hofmann, Int. J. Mod. Phys. A \textbf{20}, 4123 
(2005), Erratum-ibid. A \textbf{21}, 6515 (2006). [arXiv:hep-th/0504064]\\ 
R. Hofmann, arXiv:0710.0962 [hep-th]. 

\bibitem{SHG2006I}
M. Schwarz, R. Hofmann, and F. Giacosa, Int. J. Mod. Phys. A {\bf 22}, 1213 (2007). [arXiv:hep-th/0603078] 

\bibitem{SHG2006II}
M. Schwarz, R. Hofmann, and F. Giacosa, JHEP {\bf 0702}:091 (2007). [arXiv:hep-ph/0603174]

\bibitem{LH2008}
J. Ludescher and R. Hofmann, Ann. Phys. (Berlin) {\bf 18}, 271 (2009). [arXiv:0806.0972 [hep-th]].  

\bibitem{PDG}
http://pdg.lbl.gov/2007/listings/s000.pdf. 

\bibitem{LKGH2008}
J. Ludescher, J. Keller, F. Giacosa, and R. Hofmann, arXiv:0812.1858 [hep-th].  

\bibitem{GH2005}
F. Giacosa and R. Hofmann, Eur. Phys. J. C {\bf 50}, 635 (2007). [hep-th/0512184]

\end{thebibliography}
\end{document}